\pdfoutput=1

\documentclass[11pt]{article}

\usepackage[]{acl}

\usepackage{times}
\usepackage{latexsym}

\usepackage[T1]{fontenc}

\usepackage[utf8]{inputenc}

\usepackage{microtype}

\usepackage{inconsolata}
\usepackage{tabularx}
\usepackage{booktabs}
\usepackage{amsmath}
\usepackage{pifont}
\usepackage{xcolor}
\definecolor{green1}{RGB}{34,139,34}
\definecolor{red1}{RGB}{139,0,0}
\usepackage{enumerate}
\usepackage{multirow}
\usepackage{tablefootnote}
\usepackage{tikz}
\usepackage{amssymb}
\usepackage{caption,subcaption}
\usepackage{dsfont}
\usepackage{relsize}
\usepackage{listings}
\usepackage{algorithm}
\usepackage{algpseudocode}
\usepackage{paralist}
\usepackage{fancyvrb}
\makeatletter
\renewcommand{\ALG@beginalgorithmic}{\small}
\makeatother
\title{Numbers Matter! Bringing Quantity-awareness to Retrieval Systems}

\author{Satya Almasian, Milena Bruseva \and Michael Gertz \\
  Institute of Computer Science, Heidelberg University, Germany \\
  \texttt{\{lastname\}@informatik.uni-heidelberg.de}}

\begin{document}
\maketitle
\begin{abstract}
  	Quantitative information plays a crucial role in understanding and interpreting the content of documents. 
Many user queries contain quantities and cannot be resolved without understanding their semantics, e.g., ``car that costs less than $\$10$k''. 
Yet, modern search engines apply the same ranking mechanisms for both words and quantities, overlooking magnitude and unit information.
In this paper, we introduce two quantity-aware ranking techniques designed to rank both the quantity and textual content either jointly or independently.
These techniques incorporate quantity information in available retrieval systems and can address 
queries with numerical conditions \emph{equal}, \emph{greater than}, and \emph{less than}. 
To evaluate the effectiveness of our proposed models, we introduce two novel quantity-aware benchmark datasets in the domains of finance and medicine and compare our method against various lexical and neural models.
The code and data are available under \url{https://github.com/satya77/QuantityAwareRankers}. 

\end{abstract}

\section{Introduction}
\label{sec:intro}
Despite advances in semantic search and sophisticated neural network architectures, handling quantitative information in text remains challenging.
Specifically hard are quantity-centric queries in which the query contains a quantity and a numerical condition, e.g., ``BMW with more than 530hp''.
The reason for this is that IR systems are not aware of numbers and their semantics, such as proximity, in particular in combination with units. 
Numbers and units are treated in the same way as any other text token that is subject to subsequent processing, e.g., indexing or embedding. 
What complicates treating numbers and units in a proper way is that these objects can also have different surface forms (e.g., 6k vs 6,000 and mph vs miles per hour) and require standardization~\cite{DBLP:journals/debu/Weikum20}.  
While there are approaches that specifically focus on numbers in text, e.g., extracting quantities for entities~\cite{DBLP:conf/semweb/HoIPBW19,DBLP:conf/wsdm/LiFLLZ21}, linking quantities in tables \cite{DBLP:conf/icde/IbrahimRWZ19}, or numerical reasoning~\cite{DBLP:conf/emnlp/RanLLZL19}, they are tailored to specific tasks and not semantic search in general. 
This applies to neural models supporting Information Retrieval (IR), which are trained on general-purpose data without the focus on the quantity semantics.
Language Models (LM), forming the basis for neural models, exhibit a limited understanding of number scales and proximity~\cite{DBLP:conf/emnlp/WallaceWLSG19}.
Despite recent work on numerical language models~\cite{DBLP:conf/naacl/SpokoynyLJB22,DBLP:journals/corr/abs-2109-03137}, these architectures are very specific and require changes in the architecture of popularly used language models in IR, which indicates an expensive pre-training.
Moreover, the lack of accessible quantity-centric benchmarks for training or comparing systems exacerbates the issue.

\noindent
In this paper, we present two strategies to enhance the quantity understanding of current IR systems. 
We aim for a general-purpose model that is not specific to quantity ranking but is also capable of textual ranking. 
The two approaches differ in their integration of quantity ranking with textual ranking. 
The first employs a disjoint combination, while the second focuses on the joint ranking of quantities within the context of textual content.
The disjoint approach is an unsupervised and heuristic model utilizing an index structure, compatible with various lexical and semantic IR systems.
Due to the independent handling, the connection between quantities and surrounding text is somewhat lost. 
Therefore, for joint ranking, we aim to learn quantity-aware document and query representations through task-specific fine-tuning of neural IR models.
Additionally, we introduce two novel benchmark datasets for quantity-centric ranking, focusing on queries involving numerical conditions in finance and medicine. We evaluate our systems against various lexical and neural models and show significant improvements over the baselines.

\section{Related Work}
\label{sec:related}
\vspace*{-0.2cm}
\noindent Related work for quantity-centric search is limited. \cite{DBLP:conf/wsdm/HoPKBW20,DBLP:conf/semweb/HoIPBW19} focus on quantity search for named entities, using a deep neural network for extracting quantity-centric tuples from text and query and matching based on context similarity.  
Their pipeline involves semantic role labeling and named entity extraction, both resource-intensive and reliant on sparsely available annotated data for quantities. 
Further, focusing on named entities limits the applicability to real-world scenarios.

\noindent
QFinder~\cite{DBLP:conf/sigir/AlmasianB022} integrates numerical and lexical indexes to enhance numerical understanding in a lexical IR system. 
Our disjoint model utilizes QFinder's heuristic ranking function, but we extend their approach to include neural models and go beyond the limited query language, allowing users to provide queries in plain text. 
MQSearch~\cite{DBLP:conf/sigir/MaiyaVW15} extracts quantities with a set of regular expressions to create a rule-based system for finding documents containing certain keywords and ranges of values.
Loosely related to IR, \cite{DBLP:conf/sigir/Rybinski0KPJHTH23} and \cite{DBLP:conf/wsdm/LiFLLZ21} perform numerical summarization on unstructured text in the form of plots and graphs.

\noindent
In the area of databases, there has been some work focusing on building numerical indices for queries that contain numerical restrictions~\cite{DBLP:conf/sigir/MaiyaVW15,DBLP:journals/im/FontouraLQZ07,DBLP:journals/tkde/AgrawalS03}.  However, the main focus of such systems is the efficiency of the index structure and filtering out irrelevant numbers from the results with hard constraints rather than ranking.

\noindent
Unlike quantity-aware IR, investigating numeracy in LMs is well-established. \cite{DBLP:conf/emnlp/WallaceWLSG19} are among the first to highlight the limitations of embedding models when handling numbers.
Subsequent efforts have led to dedicated embeddings and LMs for understanding scales, basic arithmetic, and numerical common sense knowledge~\cite{DBLP:conf/naacl/SpokoynyLJB22,DBLP:journals/corr/abs-2109-03137,DBLP:conf/naacl/ThawaniPIS21,DBLP:conf/emnlp/SundararamanSSW20,DBLP:conf/emnlp/JiangNGCZST20,DBLP:journals/corr/abs-2102-13019,DBLP:conf/acl/RiedelS18}. 
These LMs are specific to numeracy and not IR in general. 
While using them can enhance performance, we focus on improving quantity understanding in current IR models without architectural changes or training a LM from scratch.




\section{Quantity-aware Model}
\vspace*{-0.2cm}
\noindent
A quantity-centric query contains a numerical condition, a value, and a unit, e.g., ``iPhone XS with price under \$$1500$ ''. 
Queries like ``What is the price of iPhone XS?'' are not considered quantity-centric as they do not require an understanding of scales and units.
In the following, we assume a document collection where each document consists of a sequence of sentences. 
Following previous work \cite{DBLP:conf/semweb/HoIPBW19,DBLP:conf/sigir/AlmasianB022}, we focus on sentences as retrieval units.
A sentence $s_i:=(T_i,Q_i)$ is a sequence of tokens $T_i=(t_1,...,t_l)$ and quantities $Q_i=(q_1,...,q_k)$, where a quantity $q_i=(u_i,v_i)$ is a tuple of a unit $u_i$ and a value $v_i$.  
A quantity query is denoted by
$X = (T_x,c,q_x)$, where $T_x=(t_{x_{1}},..,t_{x_{n}})$ are the search
terms related to the query quantity $q_x=(u_x,v_x)$. 
$c \in \{=, <, >\}$ represents a numerical condition, defining \emph{equal}, \emph{less than}, and \emph{greater than} conditions. 
\emph{Less than} and \emph{greater than}  indicate open bounds with values strictly less or greater than the query value. 
The \emph{equal} condition pertains to values strictly equal to a query value.
The relevance, $r(s_i|X)$, of sentence $s_i$ to the query $X$ is given in Eq~\ref{eq:quant-prob}. 
The similarity function $sim_c$ is dependent on the query condition $c$, where $\tau$ is a generic function that maps a query and a document to their representations. 
Here, we explore different ways to define $\tau$, which can be an embedding vector or a heuristic scoring function. 
\vspace{-0.1cm}
 \begin{equation}
 \label{eq:quant-prob}
r(s_i|X) \sim sim_c(\tau(T_x,q_x),\tau(T_i,Q_i))
\end{equation}
\noindent
 We begin with a disjoint quantity-ranking method. 
 Leveraging heuristic and supervised functions from \cite{DBLP:conf/sigir/AlmasianB022}, we extend this approach to neural models. 
 We point out the limitation of the disjoint ranking and propose a quantity-centric fine-tuning paradigm for neural IR systems for the joint ranking of quantity and textual content.
\vspace*{-0.2cm}
\subsection{Quantity Extraction} 
To facilitate both approaches, a prerequisite is a quantity extractor capable of identifying values ($v$), units ($u$), numerical conditions ($c$), and concepts ($cn$) associated with quantities. 
Concepts represent objects or events that numerical values refer to. 
For instance, in the sentence ``The iPhone 11 has 64GB of storage'', the concept is ``iPhone 11 storage''.
For this purpose, we use the Comprehensive Quantity Extractor (CQE) framework~\cite{DBLP:conf/emnlp/AlmasianKG023}. 
However, this module can be substituted with any alternative good quantity extractor.

\subsection{Disjoint Quantity Ranking}
The disjoint model is based on the separation of quantity and term ranking. 
We assume that the textual relevance of a sentence to query terms is independent of the proximity of query value and sentence values under the query condition.
Then, the relevance of a sentence can be the summation of (1) textual similarity, and (2) quantity proximity under query condition, given in Eq~\ref{eq:un-quant-prob}. 
Note that here, $sim$ computes the similarity of search terms to a sentence independent of $sim_c$, which computes the quantity proximities given query condition $c$.  
$\tau$ and $\tau^{'}$ signify that representations for query and document are not necessarily created from the same model. 
If the query is not quantity-centric, by removing the quantity score $sim_c$, the models fall back to term scoring.
 \begin{equation}
 \label{eq:un-quant-prob}
 \small
r(s_i|X) \sim sim(\tau(T_x),\tau(T_i)) + sim_c(\tau^{'}(q_x),\tau^{'}(Q_i)) 
\end{equation}
In the following, we describe the computation of (1) term ($sim(\tau(T_x),\tau(T_i))$), and  (2) quantity ($sim_c(\tau^{'}(q_x),\tau^{'}(Q_i)$) scorings. The general pipeline is depicted in App.~\ref{pipeline}.

\subsubsection{Quantity Scoring}
Using a quantity index containing explicit information about values and units in normalized form, we use heuristic functions to compute the proximity of query and sentence values based on different numerical conditions.

\noindent
\textbf{Index creation:}
Documents are split into sentences that are processed independently by CQE.
CQE outputs standardized values, e.g., \$300 million is converted to \$300,000,000 and normalized units, e.g., kilometer per hour and km/h are mapped to the same unit. 
A quantity index with unit/value pairs as keys is built from this output and resembles a lexical index.
Each unique unit/value pair points to a list of sentences it occurs. 

\noindent
\textbf{Scoring functions:}
$sim_c(\tau(q_x),\tau(Q_i))$ is estimated by a scoring function $qs$ that ranks the value in a sentence based on the value in the query given a numerical condition, where higher values indicate higher relevance. 
$qs$ is dependent on the numerical condition, resulting in different scores for the same values under different conditions. 
The quantity score only matters if the units match,
otherwise, the values are not comparable and refer to different aspects of an object, e.g.,
the horsepower of a car is different from the km/h it reaches.
$qs$ is formulated in Eq~\ref{eq:QuantScore}.
The indicator function $\mathds{1}_{u_i}(u_x)$ enforces the match between the units of the query and the sentence, and $\Phi_{c}$ is the condition-dependent scoring function.
To obtain a value between 0 and 1, the score is normalized by the number of quantities $\left| Q_i\right| $ in $s_i$.
For brevity, from now on we refer to $qs(s,c,X)$ simply as $qs$.
\begin{equation}
\label{eq:QuantScore}
qs(s_i,c,X) :=\frac{1}{\left| Q_i\right|  } \sum^{\left| Q_i\right|  }_{i=1}\mathds{1}_{u_i}(u_x)\Phi_{c} (v_{x},v_{i})
\end{equation}
\noindent
$\Phi_{c}$ refers to one of the three heuristic functions, one for each numerical condition (\emph{equal}, \emph{less than}, \emph{greater than}), adapted from \cite{DBLP:conf/sigir/AlmasianB022}. The study in \cite{DBLP:conf/sigir/AlmasianB022} explores various $\Phi$ functions and their implications for sorting of results (refer to App.~\ref{sorting}). 
Simply by changing $\Phi$s, results can be rearranged, independent of the training data.
Nonetheless, for the evaluation of our model against other baselines, we focus only on the most intuitive variant, which ranks quantities with values closer to the query value in descending order.
The $\Phi$s are defined in Eq.~\ref{eq:phis}. $v_{x}$ is the query value, and $v_i$ is the sentence value. 

\begin{equation} \label{eq:phis}
\begin{gathered}
    \Phi_{=} (v_{x},v_{i})=: exp(-|v_{x}-v_{i}|)
\\
\Phi_{>}  (v_{x},v_{i})=:
\begin{cases}
\mathlarger{v_{x}} \slash  {v_{i}} & \mathsmaller{v_{x}>v_{i}}\\0 &  \mathsmaller{else}
\end{cases}\\ 
\Phi_{<}  (v_{x},v_{i})=:
\begin{cases}
\mathlarger{{v_{i}} \slash {v_{x}}} &  \mathsmaller{v_{x}<v_{i}} \\
0&\mathsmaller{else}
\end{cases}  
  \end{gathered}
\end{equation}
$\Phi_{=}$ assesses the proximity of $v_{x}$ to $v_i$ by employing the exponential decay of their difference. The resulting score ranges between $0$ and $1$, with larger absolute differences yielding lower scores.\

\noindent
The scoring functions $\Phi_{<}$ and $\Phi_{>}$ determine numerical proximity based on the ratio of the query value $v_{x}$ to the sentence value $v_i$, resulting in a score between $0$ and $1$. This ratio, independent of magnitude, yields higher scores for closer values.

\subsubsection{Term Scoring}
Term scoring, $sim(\tau(T_x),\tau(T_i))$, can come from any lexical or semantic ranker, requiring only normalized scores.
Yet, IR systems typically do not normalize their scores, as it has no influence on the final ranking.
Here, we discuss ways to normalize scores of lexical and semantic systems and combine them with $qs$.
For a lexical model, we use BM25~\cite{robertson2009probabilistic}, and for dense and sparse neural rankers, ColBERT~\cite{DBLP:conf/sigir/KhattabZ20} and SPLADE~\cite{DBLP:conf/sigir/FormalPC21} are employed.

\noindent
\textbf{Lexical model:}
Following \cite{DBLP:conf/sigir/AlmasianB022}, we combine $qs$ with the BM25 score.
The combined score, represented in Eq~\ref{eq:QBM}, is constrained to sentences containing the search terms, as indicated by $ \mathds{1}_{T_x}(s_i)$.
The parameter $ \alpha $ controls the influence of the quantity scoring, falling back to pure term-based scoring when $\alpha$ is zero. 
The BM25$(s_i, T_x)$ score is normalized per query by dividing each sentence's score by the maximum BM25 score retrieved for the specified search terms $\max_X=max_{s \in S}(\text{\small{BM25}}(s,T_x))$.
 \begin{equation}
 \label{eq:QBM}
 \begin{gathered}
\text{\small{QBM25}}(s_i,c,X):=\  \frac{\text{\small{BM25}}(s_i,T_x)}{\max_X}\  +   \ \alpha  \mathds{1}_{T_x}(s_i) qs
\end{gathered}
\end{equation}

\noindent
\textbf{Neural dense model:}
Representing a dense neural model, ColBERT is selected for the term scoring. 
This choice is due to the same model being used for joint quantity ranking, where token-level interactions are crucial.
A contextualized term score is computed with the similarity computation between token embeddings of query and sentence, as in Eq~\ref{eq:colbert}.
ColBERT utilizes two BERT~\cite{DBLP:conf/naacl/DevlinCLT19} encoders for query ($BERT(T_x)$) and document ($BERT(s_i)$) (sentence), where each encoder outputs a list of token embeddings.
 \begin{equation}
\label{eq:colbert}
\small
 \begin{gathered}
\text{\small{ColBERT}}(s_i,T_x)= \\
\sum_{k \in [|\text{\tiny{BERT}}(T_x)|]}{max_{j \in [|\text{\tiny{BERT}]}(s_i)|} \text{\small{BERT}}(T_x)_{k} \cdot\text{\small{BERT}}(s_i)_{j}}
\end{gathered}
\end{equation}
ColBERT score comes from the MaxSim operation between the token embeddings of query and sentence.
MaxSim calculates an unbounded score for the maximum cosine similarity between the token embeddings. 
To normalize this score, we need the maximum score. 
However, calculating the maximum score for the entire collection is impractical. 
For ranking, ColBERT leverages the pruning-friendly nature of the MaxSim in an approximate nearest neighbor search~\cite{johnson2019billion} to return the top-k most relevant candidate sentences $S_k$.
We compute the maximum score based on these candidate sentences $max_{X} = max_{s \in S_k}(\text{\small{ColBERT}})$ to normalize the score between 0 and 1.
$qs$ is then exclusively applied to the top-k candidates, serving as a second-stage re-ranker for numerical proximity. 
The final score is defined in Eq~\ref{eq:qcolbert}, where $\alpha$ again controls the impact of quantity scoring.
 \begin{equation}
\label{eq:qcolbert}
 \begin{gathered}
\text{\small{QColBERT}}(s_i,c,X):=
 \frac{\text{\small{ColBERT}}(s_i,T_x)}{\max_X} + \alpha \cdot qs
\end{gathered}
\end{equation}
Note that $qs$ only affects the top-k sentences.
We also present a neural sparse model, where $qs$ is integrated into the entire ranking.

\noindent
\textbf{Neural sparse model:}
The SPLADE model extends the document and query terms and uses an inverted index for sparse dot products, allowing for end-to-end integration with the quantity scoring. 
Inside the index, instead of term frequencies, term importance weights are computed by SPLADE.
For each sentence and query, the BERT embeddings are passed through a ReLU non-linearity and log function to produce a sparse vector over the entire vocabulary, where the values of this vector are the term importance scores.
Then the relevance of the query to a sentence is based on the sparse dot product of this vector, as shown in Eq~\ref{eq:splade_score}.
 \begin{equation}
\label{eq:splade_score}
\small
 \begin{gathered}
\text{\small{SPLADE}}(s_i,T_x) := \\
\log (\mathbf{1}+\text{\small{ReLU(}}\text{\small{BERT}}(s_i))) \cdot
 \log (\mathbf{1}+\text{\small{ReLU(}}\text{\small{BERT}}(T_x)))
\end{gathered}
\end{equation}
We normalize the SPLADE score by the maximum score for a given query, $\max_X=max_{s_i \in S}(\text{\small{SPLADE}}(s_i,T_x))$, as defined in Eq~\ref{eq:QSPLADE}.
For higher precision, the quantity score is only added to sentences where there is a match between the expanded query terms and documents, denoted by the indicator
function $\mathds{1}$.
 \begin{equation}
 \label{eq:QSPLADE}
 \small
  \begin{gathered}
\text{\small{QSPLADE}}(s_i,c,X):=  \frac{\text{\small{SPLADE}}(s_i,T_x)}{\max_X} + \alpha\mathds{1}(s_i) qs 
\end{gathered}
\end{equation}
\subsection{Joint Textual and Quantity Ranking}
The independence assumption between the relevance of quantities and terms can be problematic.
Consider the query ``iPhone XR below \texteuro 200''. 
In a disjoint ranking, the following sentences can receive an inappropriately high score. 

\noindent
1)\emph{The price of an iPhone XR reached \texteuro 236.50, whereas Samsung A14 is \texteuro 132.00.} 
This sentence has multiple quantities and the numerical condition is satisfied for a value unrelated to ``iPhone XR''.  

\noindent
2)  \emph{Older iPhones, including iPhone XR have dropped in price with iPhone 8 to \texteuro 152.94.} Here, ``iPhone XR'' has no associated quantity.

\noindent
These cases are due to a lack of correct association between concept and quantity. 
We refer to this as \emph{quantity-concept mismatch}.
To address this, we need to rank sentences based on quantities in context. 
Transformer-based models inherently capture token inter-dependencies across the entire context.
However, current benchmarks lack quantity-centric data.
Therefore, it remains unclear whether the deficiency in quantity understanding is due to the absence of task-specific training data or if the current architectures hinder numerical comparisons. 
To investigate this, we propose a data generation approach to address
following problems.

\noindent
First is the inability to perform value comparisons given numerical conditions. In the example above, the models ignore the \emph{less than} condition and focus on the semantic similarity of query text and sentence.
Second, the semantic similarity of units is not well-defined. In the example above, results with ``dollar'' and other currencies receive high scores due to the context similarity of the units. Refer to App.~\ref{semantic_search} for a detailed discussion.

\noindent
Our data generation paradigm is designed to enhance \emph{value comparisons} and understanding of \emph{unit surface forms}, by generating contrastive positive and negative sentences through data augmentation.
Data augmentation, widely used in computer vision, has also found applications in NLP tasks~\citep{DBLP:conf/acl/SennrichHB16}.
The GENBERT model~\citep{DBLP:conf/acl/GevaGB20} is a relevant example, which employs templates for generating pre-training data to enhance numerical reasoning in question-answering systems without specialized architectures.

\noindent
Similar to GENBERT, we fine-tune neural IR models used in the disjoint setting, ColBERT and SPLADE, on synthetic data for quantity-centric IR~\footnote{Given that we are perturbing values and units in a sentence, one might alternatively call this \emph{data perturbation}.}.
The data generation pipeline has three stages described in the following: \emph{quantity extraction}, \emph{query generation}, and \emph{sample generation}.

\subsubsection{Quantity Extraction}\label{index-concept-unit}
The documents are split into sentences and fed to CQE to extract quantities and concepts.  
The corpus is then transformed into an index-like structure based on concepts and units. We refer to this structure as \emph{concept/unit index}.
The keys of the index are concept/unit pairs that point to a list of values associated with the pair and a list of respective sentences they occur in.
The list of values can be viewed as the distribution of values for a concept under a specific unit. 
An example entry is shown in App.~\ref{con-unit-index}.
We utilize this index structure in the subsequent steps for query and sample generation.

\subsubsection{Query Generation}
For each concept/unit pair, three queries, one for each condition, are created with the template

\begin{Verbatim}[baselinestretch=0.85,fontshape=sl]
query = {concept} {numerical_condition} 
{unit_before}{value}{unit_after}.
\end{Verbatim}
The variables enclosed in the brackets are populated during query generation.  
These steps are shown in the algorithm in App.~\ref{query_generation}. In the following, we describe how each placeholder is filled. 

\noindent
\textbf{Unit:}
A surface form of the query unit is chosen randomly from a dictionary of unit surfaces provided by CQE, e.g., ``\texteuro'' is a surface of the unit ``euro''.
 \texttt{unit\_before} and \texttt{unit\_after} account for symbols appearing before, e.g., ``\texteuro''  and abbreviations after a value, e.g., ``EUR'', respectively.

\noindent
\textbf{Value:}
For sample generation, sentences containing values meeting the query condition are crucial. 
Therefore, selecting query values with enough supporting sentences is vital. 
We propose the following strategy, based on the value distribution in the \emph{concept/unit index} for each concept/unit pair:  

\noindent
\emph{Equal query:}
Query values are chosen from the most frequent values in the index (peak of value distributions), ensuring the availability of maximum supporting sentences for a given concept/unit.

\noindent
\emph{Less and greater than queries:} For these bounds, optimal candidates are close to the mean of value distribution, such that when the numerical condition is applied more sentences fall within limits. 
Infrequent values (tail of the distribution) may have inadequate supporting sentences for the sample generation step. Refer to the App.~\ref{query_value} for examples of the value selection.

\noindent
To avoid systemic bias by focusing on the most frequent values, we generate a second set of queries for each concept/unit pair by picking the query values at random for each condition. 

\noindent
To account for variability in representation, surface forms of large values that have multiple written forms are randomly replaced with their written form. 
This takes the shape of a composite of numbers and postfixes, such as "10 million," or includes commas for digit separation, e.g., ``10,000,000''.

\noindent
\textbf{Numerical Condition:}
This is a phrase in natural language indicating a bound on a quantity, e.g., ``above'' for \emph{greater than} condition.
For this purpose, a surface-form dictionary is created, and the respective placeholder is filled with values randomly chosen from the dictionary (see App.~\ref{conditions}).

\noindent
\textbf{Concept:}
CQE identifies multi-word spans in a sentence as concepts. 
Utilizing them directly for query generation overlooks the nuances of semantic queries.
For example, in the sentence ``Disney+ charges \$6.99 a month.'', ``Disney+'' is the extracted concept. 
``Disney+'' is a streaming platform, including other media services.
Such a sentence is relevant for a lexical query with exact matches, e.g., ``Disney+ price under \$7.99 a month'', or for a semantic query, e.g., ``streaming platform price over 5 dollar/month''.
Relying exclusively on keywords in sentences poses the risk of biasing the neural models toward lexical search and away from semantic search. 
To avoid such a case, we add \emph{concept expansion}, where a large language model, namely GPT-3~\cite{DBLP:journals/corr/abs-2302-12692}, is used to generate synonyms or synsets for a given concept (see App.~\ref{concept}). 
These expansions are used to generate semantic queries. 
E.g., ``Disney+'' becomes ``Streaming platform''.
For each expanded concept, new values and unit surface forms are sampled to generate semantic queries.

\subsubsection{Sample Generation}
The input of this stage are the generated queries and the \emph{concept/unit index}.
The sample generation step creates positive and negative training samples for each quantity-centric query.
This includes positive and negative samples obtained directly from the dataset as well as additional augmented samples.
An overview of the sample generation pipeline and an algorithmic view is presented in App.~\ref{sample_generation}.
In the following, we describe each step in detail. 

\noindent
\textbf{Look-up:}
Given a query containing a \emph{(concept, unit, condition, value)}, a lookup in the \emph{concept/unit index} is performed to retrieve the sentences and the distribution of values.

\noindent
\textbf{Positive and Negative Sentences List:} 
The obtained sentences are divided into positive $s_{+}$ and negative $s_{-}$ lists, based on the numerical condition. 
$s_{+}$ contains sentences, where the values in them satisfy the query condition, and $s_{-}$ contains sentences violating the condition.

\noindent
\textbf{Original sampling:} With sample size $n$, sentences are randomly selected from $s_{+}$ as positive samples ($s_{o+}$) and from $s_{-}$ as negative samples ($s_{o-}$). Refer to App.~\ref{down_sampling} for information on the sample size.

\noindent
\textbf{Unit permutation sampling:} This method generates positive and negative samples to cover diverse unit surface forms using CQE's unit dictionary.  Positive samples contain various surface forms of the unit in the query, while negative samples include surface forms of units in the same family as the query unit, creating negatives.
\begin{itemize}
\item A positive sample, $s_{u+}$, is created by substituting the unit in a positive sentence, $s_{+}$, with other surface forms of the unit in query $u_x$.
\item A negative sample, $s_{u-}$ is created by replacing the unit in a positive sentence, $s_{+}$, with a surface form of a unit different from query unit,$u_x$, but belonging to the same family. 
The unit families are grouped based on the property they measure. For example, ``pace'', ``meter'', and ``foot'' all belong to the family of ``length''. Sampling the surface form from the same family ensures a fine distinction between unit types, even in similar contexts.
\end{itemize}
\noindent
\textbf{Value permutation sampling:} This permutation emphasizes the importance of the value comparison and numerical conditions, highlighting that sentence relevance depends on whether the sentence value satisfies the query condition or not.
\begin{itemize}
\item A positive sample, $s_{v+}$, is created by permuting the values in a negative sentence $s_{-}$, maintaining the correct concept and unit but adjusting the value to satisfy the quantity condition.
\item A negative sample, $s_{v-}$, is generated by permuting the values in a positive sentence $s_{+}$, where concept, unit, and value are all correct, to invalidate the quantity condition.
\end{itemize}
The replacement values are sampled from the values in the \emph{concept/unit index}, mirroring the underlying distribution of the relevant quantity, for to the reason for this choice, refer to App.~\ref{sampling_in_distribution}.

\noindent
\textbf{Aggregate:} The final set of positive and negative samples for each query is the union of all samples generated from the original sampling, value and unit permutation,  $s_{f+}=s_{o+}\cup s_{u+}\cup s_{v+}$ and $s_{f-}=s_{o-}\cup s_{u-}\cup s_{v-}$. 

\noindent
The models reported in the evaluation use a combination of original sampling with unit permutation and concept expansion on the query.
Value permutation did not show stable performance gains, which we attribute to the difficulty of numerical representations in dense models. 
For more discussion on this matter and the ablation study of augmentation methods refer to App.~\ref{ablation}. 

\section{Evaluation}
Given the absence of task-specific models, we assess our quantity-aware models against general domain lexical and neural models.\\
\noindent
\textbf{Lexical models} include a BM25 and a BM25$_{filter}$ variant. BM25$_{filter}$ has a separate numerical index to eliminate the results of BM25 where the query condition is not met. 
This method resembles numerical indices from databases.

\noindent
\textbf{Neural models} include the trained checkpoints of SPLADE and ColBERT as well as Cohere$_{v3}$~\footnote{\url{https://cohere.com/embeddings} DLA: 27.05.24}. 
Cohere$_{v3}$ shows that even industry-level models trained on extensive data lack quantity awareness.

\subsection{Datasets}
We introduce two English resources called FinQuant and MedQuant. 
To the best of our knowledge, these are the first quantity-centric benchmarks for retrieval.
Test queries were manually formulated using the concept/unit index, covering both lexical and semantic queries. 
Statistics for various query types are presented in Table~\ref{tab:stats_ranking}.
There is an equal number of queries for each condition, and semantic queries constitute a smaller portion due to annotation challenges.
For details on the dataset creation, see App.~\ref{dataset}.
The data is annotated by the two authors of the paper, with inter-annotator agreement computed on a subset of 20 samples per dataset. 
The Cohen's Kappa coefficient~\cite{cohen1960coefficient} is 0.83 and 0.88 for 
FinQuant and MedQuant, respectively. 
FinQuant corpus contains over 300k sentences from 473,375 news articles.
MedQuant is smaller, containing over 150k sentences from 375,580 medical documents of the TREC Medical Records~\cite{DBLP:conf/bcb/Voorhees13}.
Since the concept/unit index is used for dataset creation, CQE's performance directly affects the data quality.
While CQE is adept at handling financial data, extractions on clinical data were noisy, impacting performance comparisons later on. 
However, we find it important to report results on both datasets, making the reader aware of the lower quality of MedQuant.
\begin{table}[t]
\caption{Query types in FinQuant and MedQuant.} 
\label{tab:stats_ranking}
\setlength{\tabcolsep}{16pt} 
\renewcommand{\arraystretch}{0.9} 
\resizebox{0.5\textwidth}{!}{%
\begin{tabular}{lcc}
\toprule
   & FinQuant & MedQuant  \\
\midrule
Total queries  & 420 & 210  \\
Sentence in corpus& 306,291 & 153,252  \\
Per condition  & 140 & 70  \\
Keyword-based queries  & 300 & 120  \\
Semantic queries  & 120 & 90  \\
\bottomrule
\end{tabular}%
}
\end{table}
\vspace*{-0.2cm}
\subsection{Ranking Performance}
Table~\ref{tab:overall_ranking} shows the ranking performance of quantity-aware models, in terms of P@10, MRR@10, NDCG@10, R@100, and latency in milliseconds. 
The three models with a ``Q'' prefix indicate the disjoint and unsupervised rankers. 
Neural models with a $_{ft}$ postfix are joint models fine-tuned on synthetic data. 
Permutation re-sampling is used to test for significant improvements~\cite{riezler-maxwell-2005-pitfalls}.
Results denoted with $\dagger$ mark highly significant improvements
over the baseline models, without quantity awareness with a $p$-value < 0.01. 
All results are from single runs.
For implementation details refer to App.~\ref{implement}.

\noindent
Contrary to our initial hypothesis, disjoint rankers consistently outperform joint models across all metrics, with improvements exceeding 10 points in P@10 and over 30 points in MRR and NDCG over the base models (without the ``Q'' prefix), without requiring additional fine-tuning. 
The only drawback of the disjoint models is a minimal increase in latency, especially for QBM25 and QSPLADE, where the quantity score is added to the entire ranking.
This overhead diminishes for the top-performing model, QColBERT, where the quantity score serves as a re-ranker on the top-k candidates. 
ColBERT shows high recall on both datasets, suggesting that relevant results are within the top-k but not necessarily at the very top. 
Hence,  re-ranking with a quantity score proves beneficial.\\
\noindent
The joint models show a comparable performance boost, with metrics falling below those of the disjoint ranker but still improving from the base models. 
This validates our hypothesis that the absence of task-specific data has amplified the challenge of quantity understanding for retrieval systems.
Here, once again the ColBERT$_{ft}$ variant shows superior performance. 
We attribute the better performance of the ColBERT-based model to the fine-grained token-level interactions that allow the model to learn better associations between tokens. 
In quantity ranking, token interactions play a more significant role compared to the query and document expansions conducted by SPLADE.
This also showcases that the architecture and how the inter-token interactions are modeled matter for quantity understanding. 
Nonetheless, even after fine-tuning, understanding numerical conditions remains a challenge.
We investigate how much the fine-tuned models rely on quantities for ranking in App.~\ref{fine-tuning}.

\begin{table*}[h]
\caption{P@10, MRR@10, NDCG@10  and R@100 for on FinQuant and MedQuant. Top-2 values in each column (for each metric) are highlighted in bold.} 
\label{tab:overall_ranking}
\setlength{\tabcolsep}{8pt} 
\renewcommand{\arraystretch}{0.9} 
\resizebox{\textwidth}{!}{%
\begin{tabular}{lllllll|llll}
\toprule
 \multicolumn{1}{c}{\multirow{2}{*}{}} &\multicolumn{1}{c}{\multirow{2}{*}{Model}}& \multicolumn{1}{c}{\multirow{2}{*}{latency}}&\multicolumn{4}{c}{FinQuant}& \multicolumn{4}{c}{MedQuant} \\
\cmidrule{4-11}
  \multicolumn{1}{c}{}&\multicolumn{1}{c}{} &\multicolumn{1}{c}{(ms)} &P@10 & MRR@10 &NDCG@10 &\multicolumn{1}{c}{R@100} &P@10 & MRR@10 &NDCG@10 &R@100    \\
\midrule
\multicolumn{1}{l}{\multirow{5}{*}{baselines}}&BM25 &9& 0.06 & 0.14 & 0.09 & 0.47  & 0.04 & 0.11 & 0.07 & 0.37   \\
&BM25$_{filter}$&9 & 0.14 & 0.32 & 0.25 & 0.60 & 0.08 & 0.19 & 0.15 & 0.48  \\
&Cohere$_{v3}$ &-&0.14 & 0.22 & 0.19 & 0.27 & 0.10 & 0.17 & 0.15 & 0.25 \\
&SPLADE&26& 0.10 & 0.24 & 0.19 & 0.53 & 0.11 & 0.25 & 0.20 & 0.58  \\
&ColBERT&36& 0.15 & 0.35 & 0.27 & 0.70 & 0.12 & 0.31 & 0.24 & 0.63  \\
\midrule
\multicolumn{1}{l}{\multirow{3}{*}{joint}}&QBM25 &311& 0.21 & 0.53 & 0.41 & 0.55 & $\mathbf{0.18}$ & 0.47 & $\mathbf{0.37}$ & 0.51 \\
&QSPLADE  &319&$\mathbf{0.29}^{\dagger}$ & $\mathbf{0.67}^{\dagger}$ & $\mathbf{0.53}^{\dagger}$ & $\mathbf{0.83}^{\dagger}$ & $\mathbf{0.19}^{\dagger}$ & $\mathbf{0.52}^{\dagger}$ & $\mathbf{0.38}^{\dagger}$ & 0.70$^{\dagger}$  \\
&QColBERT &42 & $\mathbf{0.30}^{\dagger}$ & $\mathbf{0.69}^{\dagger}$ & $\mathbf{0.56}^{\dagger}$ & $\mathbf{0.87}^{\dagger}$ &  $\mathbf{0.18}^{\dagger}$ & $\mathbf{0.51}^{\dagger}$ & $\mathbf{0.37}^{\dagger}$ &$\mathbf{ 0.73}^{\dagger}$ \\
\midrule
\multicolumn{1}{l}{\multirow{2}{*}{disjoint}}&SPLADE$_{ft}$ & 26 &0.21$^{\dagger}$ & 0.51$^{\dagger}$ & 0.41$^{\dagger}$ & 0.74$^{\dagger}$ & 0.14$^{\dagger}$ & 0.37$^{\dagger}$ & 0.29$^{\dagger}$ & 0.63$^{\dagger}$ \\
&ColBERT$_{ft}$& 36$^{\dagger}$ &0.23$^{\dagger}$ & 0.55$^{\dagger}$ & 0.44$^{\dagger}$ & 0.77$^{\dagger}$ & 0.18$^{\dagger}$ & 0.44$^{\dagger}$ & 0.36$^{\dagger}$ & $\mathbf{0.72}^{\dagger}$ \\
\midrule
\multicolumn{1}{l}{\multirow{2}{*}{cross-dataset}}&SPLADE$_{out}$& 26   &-0.03 & -0.06 & -0.07 & -0.04 & -0.02 & -0.01 & -0.04 & -0.05 \\
&ColBERT$_{out}$ & 36 & -0.03 & -0.07 & -0.06 & -0.03 &-0.02 & -0.01& -0.03 & -0.02 \\
\bottomrule
\end{tabular}%
}
\end{table*}
\vspace*{-0.2cm}
\subsection{Cross-dataset Generalization}
The two lower bottom rows of Table~\ref{tab:overall_ranking} list the performance drop of joint rankers on out-of-domain data, compared to models fine-tuned on generated data from the same domain.
Each model is fine-tuned on data from the other dataset and shows a minimal performance drop, suggesting that the models learn patterns for quantity-centric queries without memorizing common queries. 
\vspace*{-0.1cm}
\subsection{Lexical vs Semantic Queries}
\begin{figure}
\centering
\begin{subfigure}[b]{\linewidth}
\includegraphics[width=\linewidth, height=0.43\linewidth]{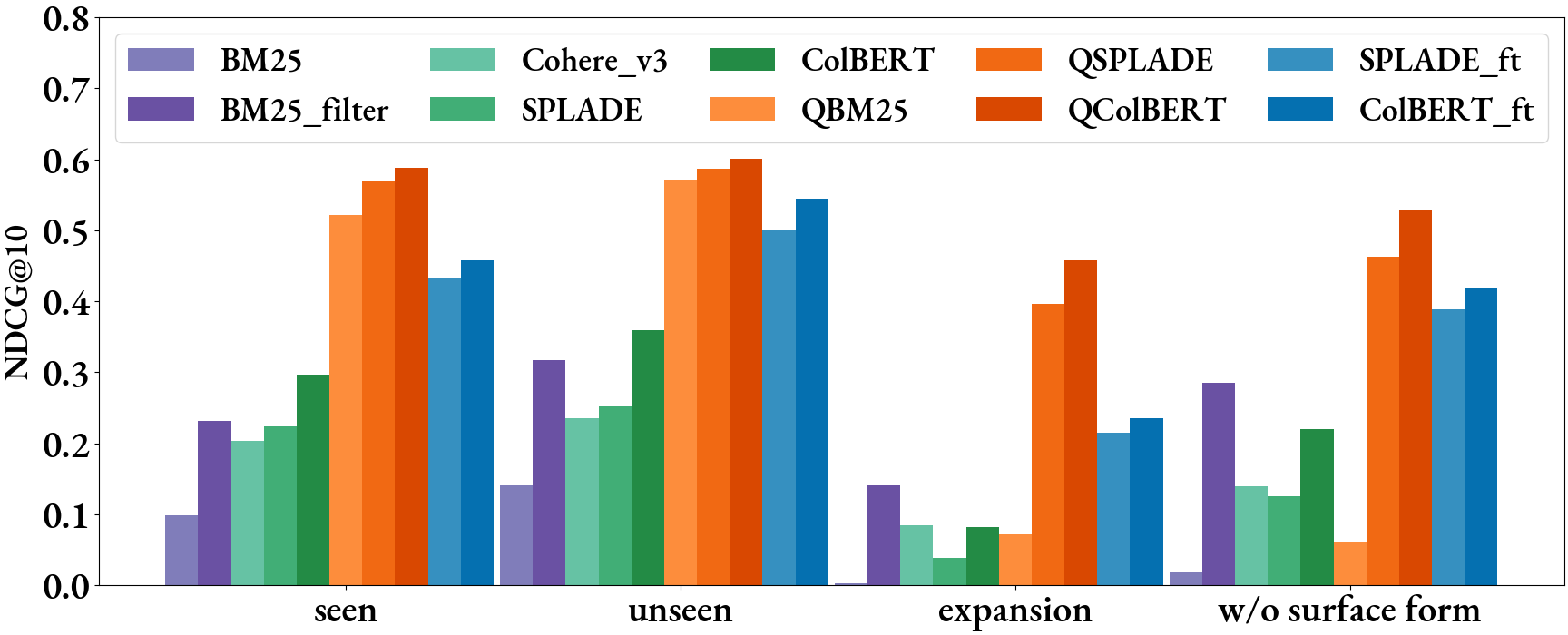}   \caption{Lexical vs semantic subsets}
   \label{fig:lexical_semantic}
\end{subfigure}

\begin{subfigure}[b]{\linewidth}
\includegraphics[width=\linewidth, height=0.43\linewidth]{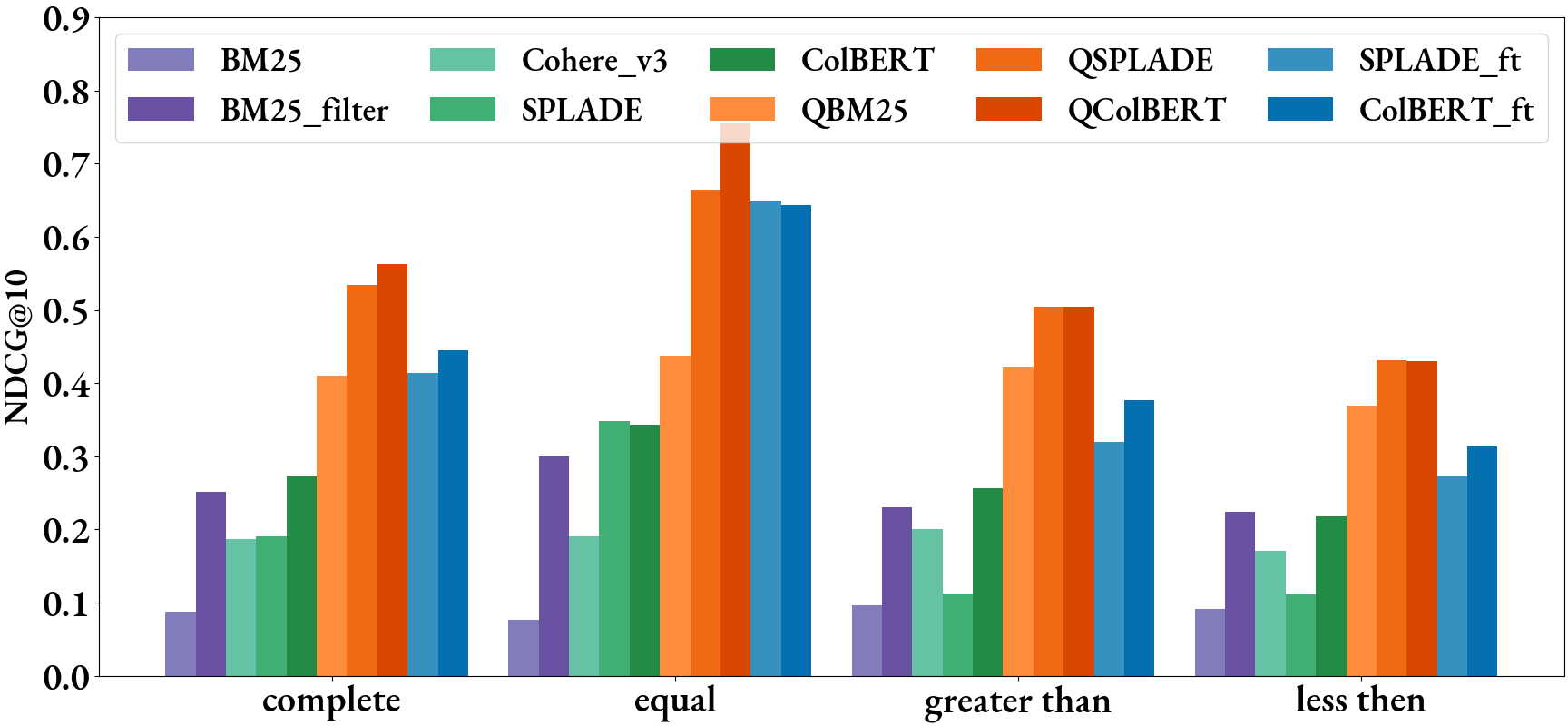}
   \caption{Subset with different numerical conditions}
   \label{fig:conditions_splits}
\end{subfigure}

\caption[]{Performance on different subsets of FinQuant. }
\vspace*{-0.3cm}
\end{figure}

Fig.~\ref{fig:lexical_semantic} shows NDCG@10 of all models on lexical and semantic subsets of the FinQuant.
The \emph{seen} and \emph{unseen} are lexical queries and \emph{expansion} and \emph{w/o surface form} represent semantic queries.  
For the details of their distinction, see App.~\ref{semnt_lex}. 
Interestingly, the disjoint ranking using dense models captures both semantic similarity and quantity understanding. 
QBM25 performs equally well in lexical queries but significantly worse on semantic ones. 
Joint rankers outperform base models, without quantity-awareness, in both lexical and semantic queries but lag behind disjoint models. 

\noindent
Fig.~\ref{fig:conditions_splits} depicts NDCG@10 of all models on different numerical conditions. 
\emph{Equal} queries are in general easier for the models as the notion of relevance in this case aligns with textual ranking. 
The performance drops almost 20 points for the bound-based conditions. 
This drop is consistent across all models, implying that the bound-based conditions are harder for models to rank.

\label{sec:eval}
\vspace*{-0.2cm}
\section{Conclusion and Ongoing Work}
\label{sec:conclusion}
In this work, we (1) examined the shortcomings prominent of IR models concerning quantity-centric queries, (2) proposed two methods, joint and disjoint quantity-aware models, to integrate quantity understanding into both classical IR models as well recent neural architectures, (3)  introduced two novel benchmark datasets containing quantity-centric queries, which to our knowledge are the first of their kind (4) demonstrated significant improvements in quantity understanding over the baselines for both proposed techniques.
Moreover, our joint and disjoint approaches enable the integration of quantity understanding without altering existing architectures and with minimal overhead in query latency.
We further highlight the strengths and weaknesses of both approaches, arguing that the choice of method should depend on the specific use case scenario.
The disjoint approach, which relies on a quantity index for ranking, where consistently outperforms joint models across various domains. 
Its unsupervised and heuristic nature also makes it more flexible than the joint rankers. 
However, despite the lower performance, the joint approach eliminates the need for an external index and the associated efficiency overhead by fine-tuning neural models on synthetic data.
Due to the lack of existing quantity-aware IR models, most of our baselines are general-purpose, but we hope that our systems can serve as baseline for future work in this direction and that our benchmarks encourage the researchers to work on this task.
In the future, we plan to explore the impact of dedicated numerical embeddings and LMs in retrieval.


\section{Limitations}
\label{sec:limitations}
In this section, we highlight the limitations of the proposed evaluation resources and the models introduced in this paper.

\noindent
\textbf{Evaluation resources:} One immediate consideration about the datasets is the relatively limited number of test queries compared to larger-scale datasets such as MSMARCO~\cite{DBLP:conf/nips/NguyenRSGTMD16}. 
This is mainly due to limited human resources and budget in an academic setting. 
Nonetheless, we argue that this number of the query is already enough to showcase certain quantity-centric capabilities. 
Another shortcoming of the data is the absence of queries for \emph{ranges}, e.g., ``iPhone with price between 500 and 800 dollars'', and \emph{negations}, ``iPhones not equal to 500 dollars''.

\noindent
\textbf{Quantity-aware models:} 
When considering neural models, one limitation is their reliance on hardware capabilities, particularly the need for GPUs to ensure efficient training, indexing, and inference.
The query latency values reported in this paper would suffer greatly if the computations were done on the CPU.
Moreover, both the synthetic data generation paradigm and the disjoint model rely on a quantity extractor. 
In the case of the disjoint model, the quality of the quantity index directly relies on the quality of value and unit extraction.
If a value and unit is not detected by the extractor it will not be considered by the scoring function. 
In the joint model, for data generation, the quantity extractor should also possess the ability to detect concepts in text, introducing the potential for additional error propagation through the system.
The performance of the quantity extractor used in this study (CQE) is not discussed here, as it is covered in detail in its respective paper. 
As for the impact of false extractions, this cannot be directly quantified because it would require a dataset with a gold standard not only for relevant passages but also for quantity extractions, which is not available. 
Furthermore, in the case of neural models, the textual and quantity ranking are intertwined, making it difficult to identify the source of the error. Nevertheless, the improvements over baseline models demonstrate that even with an imperfect quantity extractor, enhancements can still be made to existing systems.
\\
In this work, we do not discuss models that deal with \emph{ranges} and \emph{negations}.
Adding such variations to the disjoint models requires only a change in the numerical scoring function but it is more difficult for the joint setting where proper training data is required. 
For the bound-based conditions of \emph{less than} and \emph{greater than}, we considered open bounds. 
Depending on the user intent closed bounds might be more appropriate, however, similar to the optimal sorting of results, this issue does not have a single solution.

\bibliography{custom}

\appendix
\begin{figure}
\centering 
\includegraphics[width=\linewidth, height=0.48\linewidth]{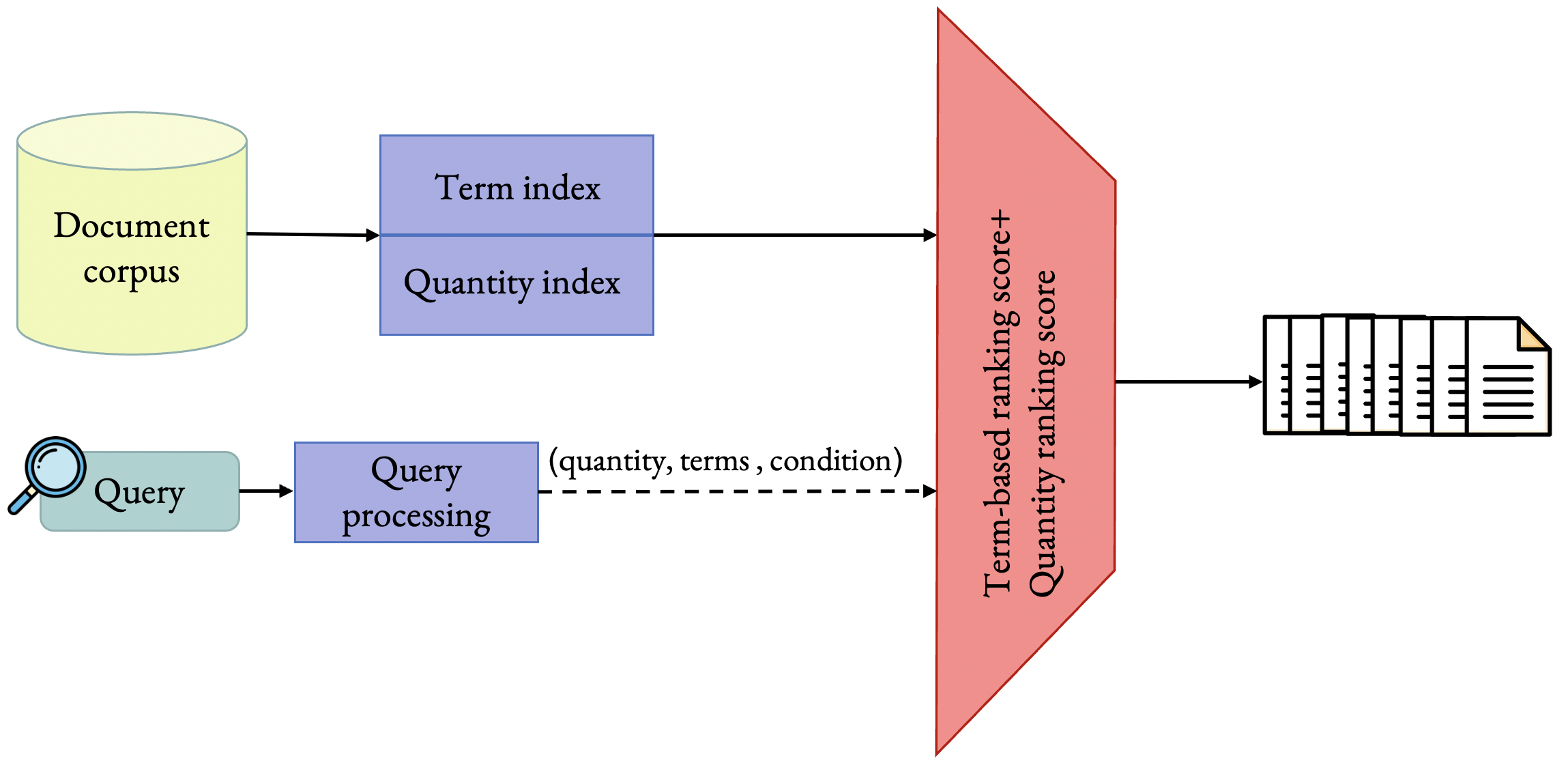}
\caption{Pipeline of the disjoint quantity-ranking approach, where a separate quantity index facilitates the computation of quantity proximity and a term-based lexical or semantic index is used to compute the similarity of the search terms to sentences. }
\label{fig:unsupervised-ranking}
\end{figure}

\begin{figure*}
\centering 
\includegraphics[width=0.8\linewidth, height=0.14\linewidth]{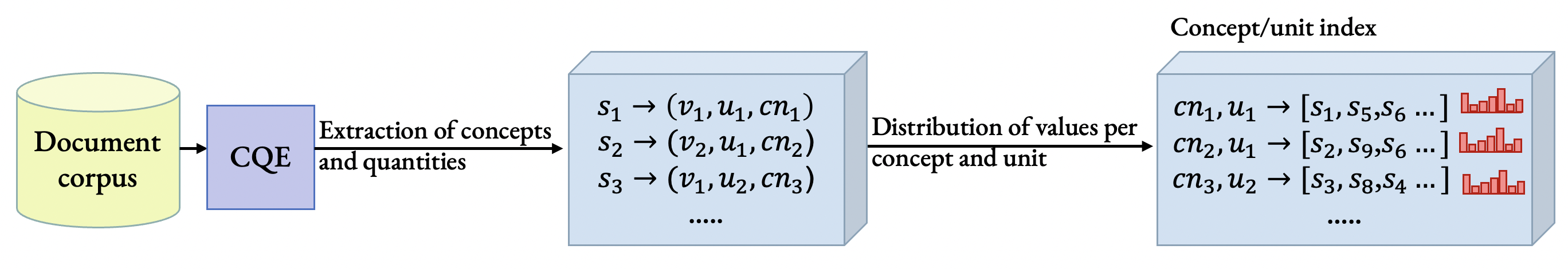}
\caption{Overview of the quantity tagging step and creation of \emph{concept/unit index} structure. }
\label{fig:quantity_tagging}
\end{figure*}

\begin{figure*}
\centering 
\includegraphics[width=0.7\linewidth]{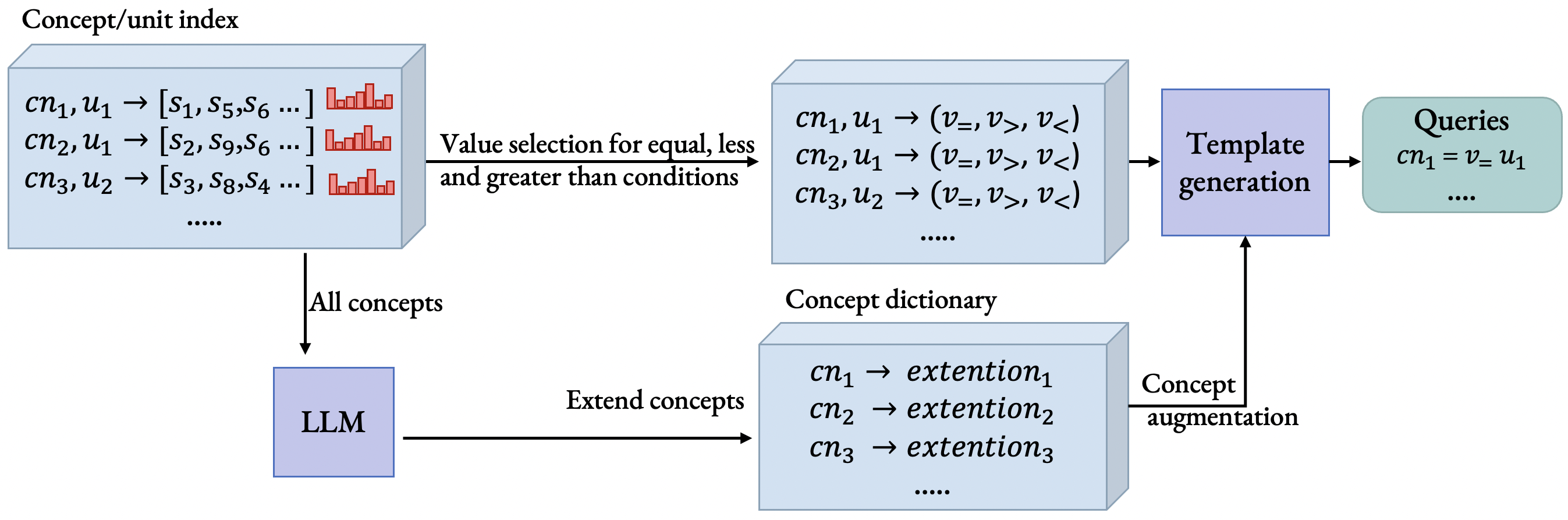}
\caption{Overview of the query generation pipeline, using \emph{concept/unit} index and a large language model for concept expansion.  }
\label{fig:query_generation}
\end{figure*}

\begin{figure}
\centering 
\includegraphics[width=\linewidth, height=0.4\linewidth]{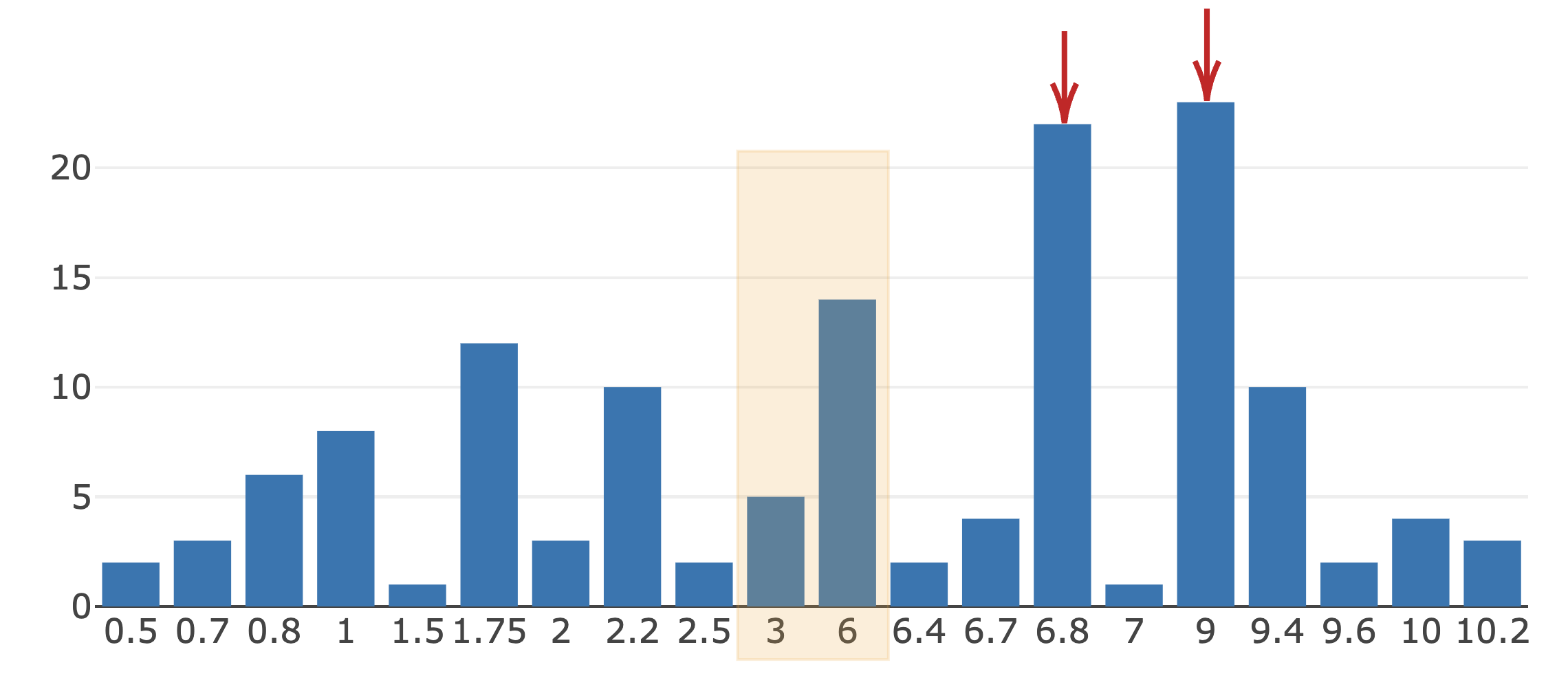}
\caption{An example of choosing query values for \emph{equal} and bound-based conditions. }
\label{fig:distribution}
\end{figure}
\section{Disjoint Quantity Ranking}
In this section, we provide additional material related to the disjoint quantity ranking model. 
\subsection{Disjoint Quantity Ranking Pipeline}\label{pipeline}
The pipeline for the disjoint quantity-ranking model is shown in Fig.~\ref{fig:unsupervised-ranking}.
The query is decomposed into quantity, search terms, and condition, using CQE or a similar package. 
The document corpus is indexed separately for terms and quantities. 
The term-based index can be a traditional lexical index or a vector database, which retrieves semantically or lexically similar sentences. 
From the retrieved sentences, the quantity index identifies values that share the same unit as the query, computing proximity based on the provided heuristic ranking functions. 
The final ranking combines scores from term-based and quantity ranking.

\subsection{Optimal Sorting}\label{sorting}
Although all the sentences that satisfy a query condition and have the correct concept and unit are
potentially relevant to the query, the order in which the result items are presented to the user can either aid or hinder the user in finding the desired result.
In term-based ranking, the optimal order of results is evident. 
However, when it comes to quantities, relevance is more subjective and the optimal sorting is dependent on the user's information needs. 
For example, a user searching for ``iPhone camera that has more than $8$ inches'' might look for a maximum value larger than $8$ inches or a display only marginally larger, both of which are valid answers. 
Presenting results in ascending or descending order based on numerical distances allows the user to identify the desired result more efficiently. 
\cite{DBLP:conf/sigir/AlmasianB022} briefly addresses this issue and explores potential alternatives for scoring functions to enable various sorting options. \\
Our disjoint approach is flexible concerning different sorting. 
By switching a scoring function, the results can be rearranged.
The joint models proposed here are not as adaptive, and rearranging the results requires additional fine-tuning based on a new preferred sorting. 

\section{Joint Quantity Ranking}
In this section, we provide additional material related to the joint quantity ranking model. 

\subsection{Concept-unit Index}\label{con-unit-index}

An example entry in the \emph{concept/unit index} from the FinQuant dataset is shown below. 
\begin{Verbatim}[baselinestretch=0.85,fontshape=sl]
{("cannabis company", "cent per share"): 
{"values":[0.9, 1.4, 17.0, 17.0, 22.0, 26.0, 
35.0, 84.0],
"sentences":['The cannabis company says
 the loss amounted to 0.9 of a cent per 
 share for the quarter ended May 31 
 compared with a loss of \$4 million or 
 1.4 cents per share a year earlier .',
'The cannabis company says its loss 
amounted to 17 cents per diluted share
 for the quarter ended Jan. 31 .',...]}}
\end{Verbatim}
Note that the repetition of values for the same concept/unit pair is stored as duplicates, such that the frequency of values is kept for the distribution, e.g., the value ``17.0'' is repeated twice as it occurs in two sentences.
The creation steps are depicted in Fig.~\ref{fig:quantity_tagging}.
The corpus is processed with CQE to extract values, units, and concepts from each sentence, where sentences sharing the same unit and concept are grouped into a list, along with values represented as a frequency distribution.

\subsection{Query Generation}\label{query_generation}
The complete query generation pipeline is depicted in Fig.~\ref{fig:query_generation}. 
The \emph{concept/unit index} is used to select values and units for numerical conditions. 
Additionally, a large language model is used to expand concepts for semantic queries.
The template generation block combines all the outputs of other blocks to formulate three queries for each concept/unit pair.
To generate queries for expanded concepts, a new query value is chosen from the value distribution, and a new set of queries is formulated. 
Additionally, we offer the query generation pseudocode in Algorithm~\ref{alg:query_generation} to make the input and output of each step clear. 
In the algorithm, $v$ refers to a value, $u$ to a unit, $c$ to a condition, and $cn$ to a concept.

\begin{algorithm}
\caption{Query Generation}\label{alg:query_generation}
\begin{algorithmic}
\Function{generate\_query}{$cn, u, c$}

 $v \gets$ get\_query\_values($cn\_unit\_dict , c$)
	
	$u\_b, u\_a \gets$ get\_unit\_surfaceform($u$) 
	
	$c \gets$ get\_condition\_surfaceforms($c$) 
	
	$query \gets conc + c + u\_b+ v + u\_a$
	
	\Return query
	
\EndFunction
\\\hrulefill
\State cn\_unit\_dict $\gets$ concept/unit index
	
\State cn\_expand\_dict $\gets$ concept expansions

\For{$(cn, u)$ in cn\_unit\_dict}: 

	\For{$cn$ in $[cn$, cn\_expand\_dict$[cn]]$}: 
	\For{$c$ in $(equal, greater, less)$}: 
	
	\textbf{\scriptsize GENEARATE\_QUERY}$(cn, u, c)$

	\EndFor
	\EndFor
\EndFor
\end{algorithmic}
\end{algorithm}

\subsection{Choosing the Right Query Value}\label{query_value}

Each entry in the \emph{concept/unit index} points to the sentences and list of values in those sentences. 
For the data augmentation to work, we require a number of positive and negative samples per query and therefore, it is important to choose the value of the query such that supporting sentences in the corpus are present. 
A hypothetical example of value distribution is shown in Fig.~\ref{fig:distribution}.
For the \emph{equal} query, the challenge is to find enough positive samples, since there is an abundance of different values in each distribution.
In FFig.~\ref{fig:distribution}, values with the highest frequency, denoted by red arrows pointing to peaks in the distribution, serve as optimal candidates for the equal condition. 
In this manner, we make sure that there are enough positive samples for the data augmentation. 
Values close to the average (highlighted in a yellow box) are chosen for the \emph{less than} and \emph{greater than} queries. 
For such queries, we avoid 
infrequent values towards the tail of the distribution, to circumvent too few positive or negative samples.

\subsection{Dictionary of Numerical Conditions}\label{conditions}
A non-comprehensive dictionary of surface forms for numerical conditions is shown Table~\ref{tab:changes_ranking}, containing multiple surface forms for each condition.
\begin{table}[h]
\caption{Numerical conditions used for query generation and their surface forms. }
\label{tab:changes_ranking}
\setlength{\tabcolsep}{5pt} 
\renewcommand{\arraystretch}{1.2} 
\resizebox{0.5\textwidth}{!}{%
\begin{tabular}{lcc}
\toprule
 Condition &   Surface forms \\
\midrule 
 Equal &  exactly, exact, equals, equals to, for, with, of, at   \\
 greater than   & greater than, more than, above, larger than, over, higher than, exceed, exceeding\\
 Less than & smaller than, below, less than, fewer than, no more than, beneath \\

\bottomrule
\end{tabular}%
}
\end{table}
\subsection{Concept Expansion}\label{concept}
For concept expansion, we use the OpenAI API~\footnote{\url{https://openai.com/} DLA:27.05.2024} and employ the \texttt{text-davinci-003} model with few-shot learning. We set the temperature to 1 to encourage creative responses. 
Since the concepts come from two distinct domains of finance and medicine, the few-shot examples vary accordingly. 
Below we specify the two prompts used for concept expansion, the result is stored in a concept expansion dictionary and utilized during query generation. 
The place-holder \texttt{\{concept\}} is replaced with a concept from the \emph{concept/unit index} that is meant to be expanded. 
 \\

\noindent
For the financial domain:
\begin{Verbatim}[baselinestretch=0.85,fontshape=sl]
Complete with the words super set or 
synonym, but do not reuse the exact 
same words, the word "Super Set" 
should not be in the response and 
response should have at least two words:

S&P 500 = stock market index
Audi = car
Oil prices = petroleum prices
unemployment rate = unemployment 
percentage
iPhone sales = phone sales
Netflix shares = stock shares
President Trump = President
iPhone 11= iphone
Hong Kong = city
stake PEXA = Property Exchange Australia 
shares

{concept} =
\end{Verbatim}

\noindent
For the medical domain:
\begin{Verbatim}[baselinestretch=0.85,fontshape=sl]
Complete with the words super set or 
synonym, but do not reuse the exact 
same words, the word "Super Set" 
should not be in the response and 
response should have at least two words:

ophthalmic solution = eye medication
Control group = treatment group
irinotecan hydrochloride = chemotherapy 
drug
monoclonal antibody = substitute 
antibodies
MRI scans = Magnetic resonance imaging
influenza H1N1 vaccine = flu vaccine
HAI antibody response = Influenza-specific 
antibody response

{concept} =
\end{Verbatim}

\subsection{When Semantic Search Backfires}\label{semantic_search}
Semantic retrieval systems consider an entire context to find relevance to a query at hand. 
Often, this aligns well with the user's expectations. 
For instance,  when searching for a  ``dark color evening dress'', any dress that can be worn as an evening gown and has a dark color would be suitable.
But as soon as the user becomes more specific like ``blue evening dress'', the embedding space could also bring a similar color like ``teal'' into the search result. 
Depending on the user's flexibility regarding the dress color, this behavior may or may not be desirable.
Such hard constraints are challenging for neural models. 
Quantity-centric queries impose hard constraints on values and units where the fuzzy matching of context might do more harm than good. 
For instance, when searching for a ``car with more than 320 hp'', if the results contain a car with ``360 brake horsepower'' instead of horsepower, the result is irrelevant. 
Both horsepower and brake horsepower are used in similar contexts but refer to different attributes. 
Horsepower measures the power generated by the engine, while brake horsepower measures how much of the power produced by the engine is sent to the wheels that make the car accelerate. 
Another common problem is with currencies. 
Given that monetary values often appear in similar contexts, it becomes challenging for neural models to differentiate between various currency units.
The same applies to hard constraints on values, where based on a given numerical condition, values outside of that bound are considered irrelevant.

\subsection{Sample Generation with Permutations}\label{sample_generation}
\begin{figure*}
\centering 
\includegraphics[width=0.7\linewidth]{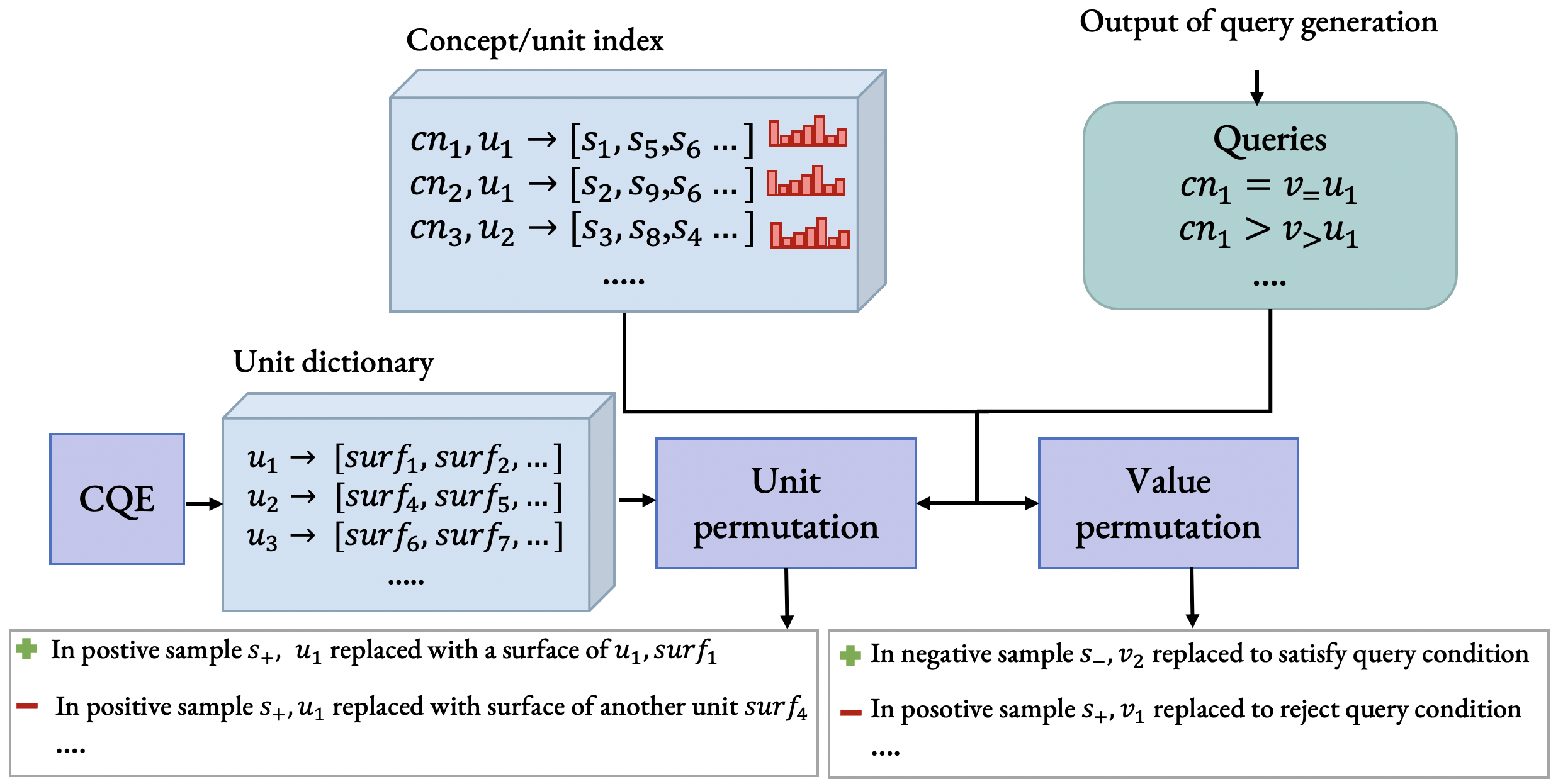}
\caption{Overview of the sample generation using value and unit permutation.   }
\label{fig:sample_generation}
\end{figure*}

An outline of the sample generation pipeline is shown in Fig.~\ref{fig:sample_generation}. 
The input of this stage is the generated queries and the \emph{concept/unit index}.
For each quantity-centric query, a list of positive and negative samples is created by applying the numerical condition on the list of sentences from the index.
The original positive and negative samples are then chosen at random from such a list. 
The same list is utilized as seed samples for data augmentation. 
Unit and value permutation are employed to generate augmented positive and hard negative samples. 
Hard negative are positive samples, where the unit or value is perturbed to violet the query condition.\\
The steps are presented in Algorithm~\ref{alg:sample_generation}.
Each sampling mechanism is encapsulated within a distinct function and the final training samples are the union of all generation mechanisms. 
In the algorithm, $v$ refers to a value, $vals$ to the list of the values of a given concept and unit, $u$ to a unit, $c$ to a condition, $cn$ to a concept, and $n$ to the sample size. 

\begin{algorithm}
\caption{Sample Generation}\label{alg:sample_generation}
\begin{algorithmic}
\Function{original\_sampling}{$s_+, s_-, n$}

\Return sample(	$s_+$,$n$), sample(	$s_-$,$n$)

\EndFunction
\\\hrulefill

\Function{unit\_permutation}{$s_+, n, u$}

	$s_{u+} \gets$ replace\_same\_unit\_surface($s_+, u$)
	
	$s_{u-} \gets$ replace\_other\_unit\_surface($s_+, u$)
	
\Return sample(	$s_{u+}$,$n$), sample(	$s_{u-}$,$n$)

\EndFunction
\\\hrulefill

\Function{v\_permuation}{$s_+, s_-, n, vals, c$}

	$s_{u+} \gets$ replace\_with\_positive\_value($s_-, v$)
	
	$s_{u-} \gets$ replace\_with\_negative\_value($s_+, v, c$)
	
\Return sample(	$s_{v+}$,$n$), sample(	$s_{v-}$,$n$)

\EndFunction
\\\hrulefill
\State conc\_unit\_dict $\gets$ concept/unit index
	
\State queries $\gets$ list of queries

\State n $\gets$ number of samples

\For{$(cn, u, c, v)$ in queries}: 

	$s,vals \gets$ conc\_unit\_dict[$(cn,u)$] 
	
	$s_{+}, s_{-} \gets$ filter\_based\_on\_condition($s, c, v$)
	
	$s_{o+}, s_{o-} \gets$ \textbf{\scriptsize ORIGINAL\_SAMPLING}$(s_+, s_-, n)$
	
	$s_{u+}, s_{u-} \gets$ \textbf{\scriptsize UNIT\_PERMUTATION}$(s_+, n, u)$
	
	$s_{v+}, s_{v-} \gets$ \textbf{\scriptsize V\_PERMUTATION}$(s_+, s_-, n, vals, v, c)$
	
	$s_{f+}=s_{o+}\cup s_{u+}\cup s_{v+}$ 
	
	$s_{f-}=s_{o-}\cup s_{u-}\cup s_{v-}$
	
\EndFor
\end{algorithmic}
\end{algorithm}

\subsection{Sampling within Distribution}\label{sampling_in_distribution}
It is crucial that the permutation values obey the original value distribution of the corpus. 
The properties of concepts are often limited to a specific range, e.g., the value ``10000'' is an unreasonable unemployment rate.
Moreover, certain values are on a discrete scale with limited options, e.g., ``RAM of a laptop'' is limited to distinct values such as 4,8, and 16.
Assigning a random number outside this range, like 10, would be unrealistic. 
Therefore, for the synthetic data to obey the rules of the real-world datasets and reflect the distribution of different properties, the permuted values are chosen from the values observed in the corpus.

\subsection{Down-sampling}\label{down_sampling}
If the number of available sentences in the positive and negative lists is smaller than the sample size, a \emph{downsampling} procedure is implemented.  When $|s_{+}|< n$ or $|s_{-}|< n$, we reduce the sample size to the smallest number of available samples.

\section{Evaluation}
In this section, we present additional evaluations and implementation details. To reproduce the results or to access the trained model checkpoints and datasets, we encourage the reader to refer to our repository.

\subsection{FinQuant and MedQuant Datasets}\label{dataset}
In this section, we give an overview of the creation of the FinQuant and MedQuant evaluation benchmarks. 
FinQuant is created from a set of news articles in the categories of economics, science, sports, and technology, collected between 2018 and 2022.
MedQuant contains TREC Medical Records~\cite{DBLP:conf/bcb/Voorhees13} on clinical trials. 
Both datasets were split into sentences and processed to eliminate boilerplate HTML and headers.
All sentences containing quantities were incorporated into the collection. 
The entire test data is manually created and tagged. 
In the following, we describe the query formulation and annotation task.\\
 
\noindent
\textbf{Query formulation: }
Given access to the \emph{concept/unit index} and the value distributions, annotators were tasked to formulate quantity-centric queries. 
During formulation, they were instructed to scan the entire index for possible synonyms for a given concept and keep track of the synonyms in a list.
For example, if one chooses ``Microsoft Surface Earbuds'' with the unit ``pound sterling'', the annotator scans the other concepts inside the  \emph{concept/unit index} that co-occur with ``pound sterling'' to detect synonyms or synsets, e.g., ``Earbuds'' and ``Microsoft headphones'' are related to the concept ``Microsoft Surface Earbuds''.
In the subsequent stage, the value distributions of all selected concepts are consolidated into one and presented to the annotator.
The annotator is then instructed to choose three values for \emph{equal}, \emph{less than}, and \emph{greater than} queries, in such a way that supporting sentences for the query are present within the value distribution. 
In the final stage, the annotator will formulate the query in natural text, e.g., ``Microsoft Surface Earbuds lower than 179 pound sterling''. 
The annotators have access to the dictionary of surface forms for units and conditions to help with the query formulation. \\

\noindent
\textbf{Candidate list generation :}
For each query, a list of relevant sentences as candidates was generated using the \emph{concept/unit index}. All sentences related to the concept and its synonyms were filtered based on the query value and condition.
The filtering is done automatically based on the query value and numerical condition to lower the effort of annotation.
We recognize that the quality of the candidate set relies directly on how effectively the quantity extractor captures associations between quantity and concepts.
We observed that although the extractions for financial data were of high quality,  in the medical domain, several quantities were overlooked. 
In both datasets, there is no guarantee that the candidate list is comprehensive and covers all relevant instances.\\

\noindent
\textbf{Annotation:}
An annotation guideline was devised for consistent annotation of ambiguous cases and is published with the dataset. 
Annotators were presented with a list of candidate sentences for each query and were tasked to mark the relevant sentences. 
The marked sentences are used as ground truth for subsequent evaluation.

\subsection{Semantic and Lexical Queries}\label{semnt_lex}
The queries from the test set are categorized into four types: \emph{seen}, \emph{unseen}, \emph{expansion}, and \emph{w/o surface form}.
The lexical queries fall under the categories of \emph{seen} and \emph{unseen}. 
For such cases, during query formulation, the annotators picked concepts from the \emph{concept/unit index} without any change in their surface form. 
The concepts from the \emph{unseen} category, were removed from the index for data generation and training of the joint neural models. 
Therefore, this subset contains lexical queries that were not seen during training. 
For example, ``YouTube channel'' is a concept in the \emph{unseen}  subset, which means all instances of ``YouTube channel'' were removed from the \emph{concept/unit index} before data generation.

\noindent 
Semantic queries contain the two subsets of \emph{expansion} and \emph{w/o surface form} and were slightly harder to formulate, thereby, fewer instances of them are present in the data. 
For \emph{expansion} queries, a concept from the lexical set was chosen to expand to one of its supersets or synonyms. 
For example, ``social media channel'' is a semantic concept from ``YouTube channel''.
These expansions were used to formulate queries that did not have a lexical match in the database and often included a superset of many concepts. 
In the case of ``social media channels'', the model should be able to retrieve other social media channels like ``Facebook'' as well as ``YouTube''.
In the case of lexical models based on BM25, the difference is evident in Fig.~\ref{fig:lexical_semantic}, where the models show great performance on \emph{seen} and \emph{unseen} subset, but if the same queries are converted to their semantic counterpart, as in \emph{expansion}, the models fail to retrieve the correct result. 
\emph{W/o surface form} are other semantic queries that were formulated independent of the lexical queries. 

\subsection{Implementation}\label{implement}
The code is implemented in Python 3.10.9 and PyTroch 1.13.1.
The general sentence splitting and text cleaning were performed with
SpaCy 3.6~\footnote{\url{https://spacy.io/} DLA: 27.05.2024}. 
As mentioned before we use the CQE
library~\footnote{\url{https://github.com/vivkaz/CQE} DLA: 27.05.2024} for
quantity extraction.
Evaluation and metrics were computed with the help of \texttt{pytrec\_eval} library~\cite{VanGysel2018pytreceval}.
In the following, we discuss the implementation details for each model separately. \\

\noindent
\textbf{BM25 models:}
We use the Okapi BM$25$ package~\footnote{\url{https://pypi.org/project/rank-bm25/} DLA: 27.05.2024} for all BM25 variants. 
The QBM25 and BM25$_{filter}$ are variations of Okapi BM$25$ designed to include a numerical index for ranking and filtering. 
The parameters of BM25 were tuned to each of our datasets separately, as presented in Table~\ref{tab:parameters}.  
The latency values are computed with plug-ins for an  Opensearch~\footnote{\url{https://opensearch.org/} DLA: 27.05.2024} instance on a desktop computer with 16GB of RAM. 
In comparison to the dense models, the lexical models do not require specific hardware architectures.

\begin{table}[h]
\caption{Hyper parameters of BM25-based models on the benchmark datasets.} 
\label{tab:parameters}
\setlength{\tabcolsep}{6pt} 
\renewcommand{\arraystretch}{1} 
\resizebox{0.5\textwidth}{!}{%
\begin{tabular}{lll}
\toprule
   & FinQuant & MedQuant  \\
\midrule
BM25  & b = 0.5, k1 = 0.5,  &  b = 0.5, k1 = 0.5  \\
BM25$_{filter}$  & b = 0.75, k1 = 1.5 & b = 0.75, k1 = 1.5   \\
QBM25  & b = 0.5, k1 = 0.5 &  b = 0.5, k1 = 0.75  \\
\bottomrule
\end{tabular}%
}
\end{table}
\noindent
\textbf{Cohere baseline:}
We used the Cohere API~\footnote{\url{https://cohere.com/} DLA: 27.05.2024} for Cohere$_{v3}$ embeddings. Query embeddings were used to encode the queries and document embeddings to encode the collection. \\

\noindent
\textbf{ColBERT models:}
\cite{DBLP:conf/sigir/KhattabZ20} supplied the trained checkpoint for the base ColBERT model. 
For fine-tuning augmented data, the model was initialized with this base checkpoint. 
The checkpoint was employed for the evaluation of both ColBERT and QColBERT. 
ColBERT$_{ft}$ was fine-tunned using the training script from the official repository~\footnote{\url{https://github.com/stanford-futuredata/ColBERT} DLA:27.05.2024}. 
The code in the repository was modified to establish an endpoint for QColBERT, incorporating a quantity index. 
We did not perform extensive hyperparameter tuning except for the learning rate and mainly used the parameters advised by the authors for both FinQuant and MedQuant datasets. 
We fine-tuned the joint ColBERT$_{ft}$ for 2 epochs, with a batch size of 256 and a learning rate of 1e-05 on a server with four A-100 GPUs and 40GB of memory. 
The evaluation and benchmarking for latency were performed on the same server, utilizing all four GPUs. 
\\

\noindent
\textbf{SPLADE models:}
SPLADE$_{ft}$ was also fine-tuned using the training script by the authors~\footnote{\url{https://github.com/naver/splade} DLA:27.05.2024}. 
The pre-trained checkpoint was acquired from HuggingFace~\footnote{\url{https://huggingface.co/naver/splade-cocondenser-ensembledistil} DLA:27.05.2024} and utilized for both the SPLADE model and QSPLADE. 
Scripts from the official repository were adjusted to add a quantity index for QSPLADE. 
Similar to ColBERT, we conducted limited hyperparameter tuning, mainly focusing on the learning rate.
We fine-tuned SPLADE$_{ft}$ for 2 epochs using a batch size of 240, a learning rate of 2e-5, and a weight decay of 0.01. 
The fine-tuning was conducted on a server with four A-100 GPUs and 40GB of memory.
The evaluation and benchmarking for latency were performed on the same server, utilizing all four GPUs. \\

\noindent 
For all disjoint rankers, QBM25, QColBERT, and QSPLADE, the quantity impact parameter of $\alpha$ is set to 1, such that the impact of term and quantity ranking are equal. \\

\noindent
\textbf{Generated data:}
Based on the combination of augmentation methods the size of training data would vary.
In all cases, we saved a small sample of 1000 queries for validation.
There were 40,732 and 20,376 concept and unit pairs considered for query generation in FinQuant and MedQuant, respectively.
If concept expansion is applied these numbers would double to account for queries on expanded concepts. 
We set the sample size $n$ to 2, meaning that for each query two positive and the negative samples were chosen from the data without augmentation. 
As a result, based on augmentation methods, additional $n=2$ samples would be added for unit and value permutation, a total of $3n$ per query.

\subsection{Effect of Fine-tuning}\label{fine-tuning}

To assess the impact of task-specific fine-tuning on the internal ranking strategy of the dense models, we evaluate two masked versions of the data.

\noindent
\emph{Mask value:} 
In this scenario, we mask all values in the collection with the \texttt{[MASK]} token before running the evaluation. This task aims to determine the extent to which the model depends on the value token for retrieving the correct sentence.

\noindent
\emph{Mask unit:} 
Here, we mask unit tokens in the collection before running the evaluation with \texttt{[MASK]} token.
This task is intended to observe the impact of unit comparison on the final ranking. 

\begin{figure}
\centering
\begin{subfigure}[b]{\linewidth}
\includegraphics[width=\linewidth, height=0.29\linewidth]{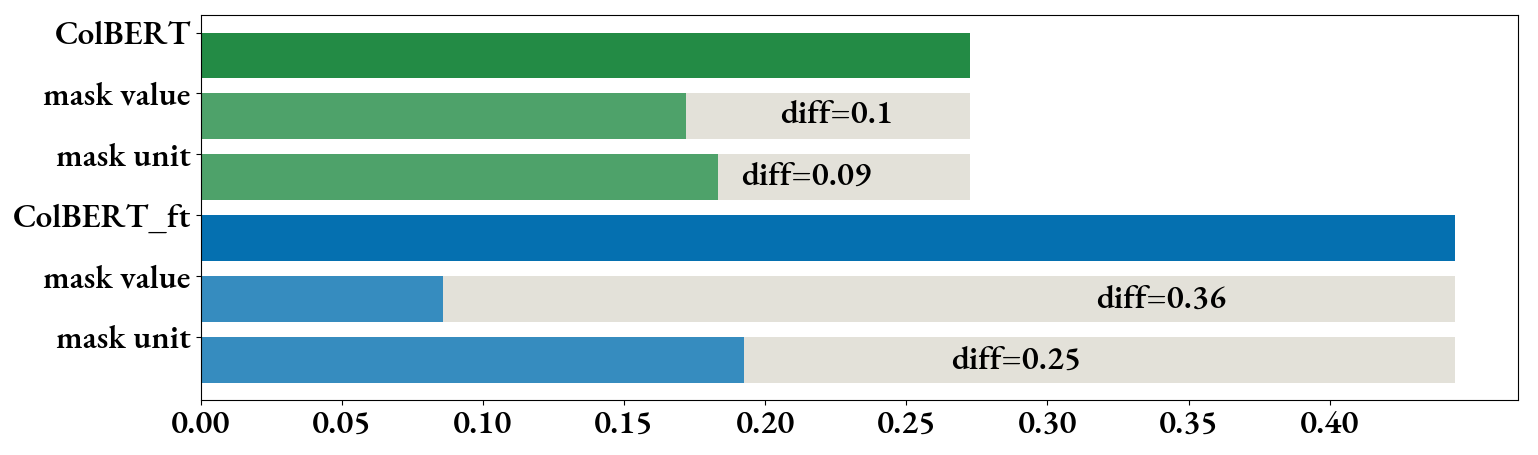}   \caption{ColBERT based models}
   \label{fig:colbert_mask} 
\end{subfigure}

\begin{subfigure}[b]{\linewidth}
\includegraphics[width=\linewidth, height=0.29\linewidth]{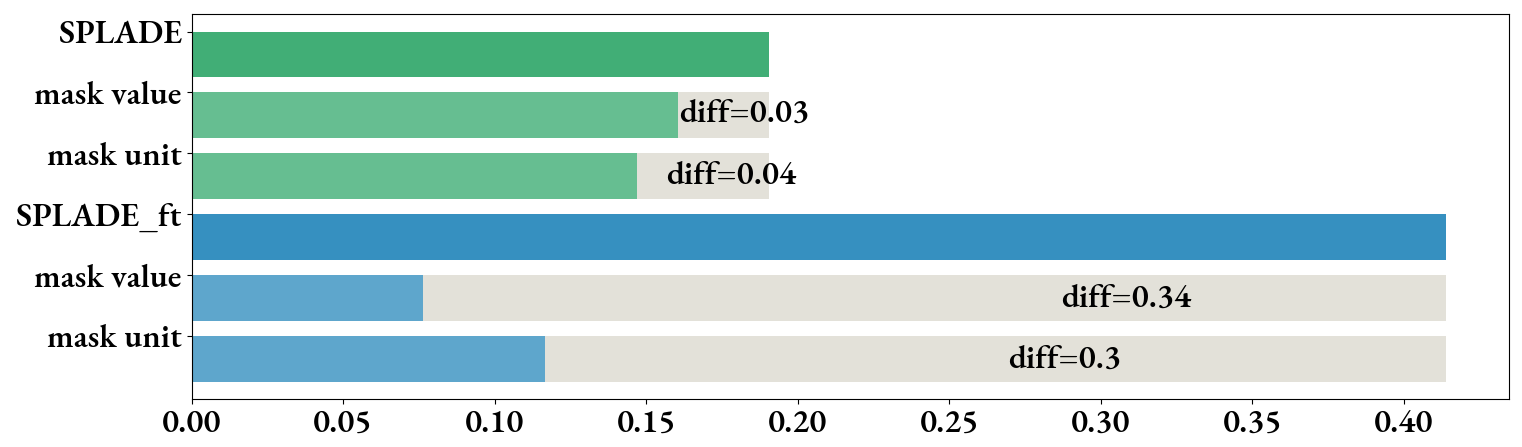}
   \caption{SPLADE based models}
   \label{fig:splade_mask}
\end{subfigure}

\caption[]{The effect of task-specific fine tuning on models attention to quantity tokens. In the masked variants either the unit or the value of the sentences in the collection is masked. }
\end{figure}

\noindent 
We compare the base version of the dense models with their fine-tuned version on the different masking of the FinQuant dataset. 
The results for the ColBERT models are shown in Fig.~\ref{fig:colbert_mask} and for SPLADE models in Fig.~\ref{fig:splade_mask}.
In both cases, the fine-tuned version exhibits a more significant drop in performance compared to the base models when quantity tokens are masked. 
This indicates that after fine-tuning, the model becomes more dependent on the quantity tokens, namely, values and units, in the text to identify the relevant sentence.

\subsection{Ablation Study on Augmentation Methods}\label{ablation}
\begin{figure}
\centering
\begin{subfigure}[b]{\linewidth}
\includegraphics[width=\linewidth, height=0.29\linewidth]{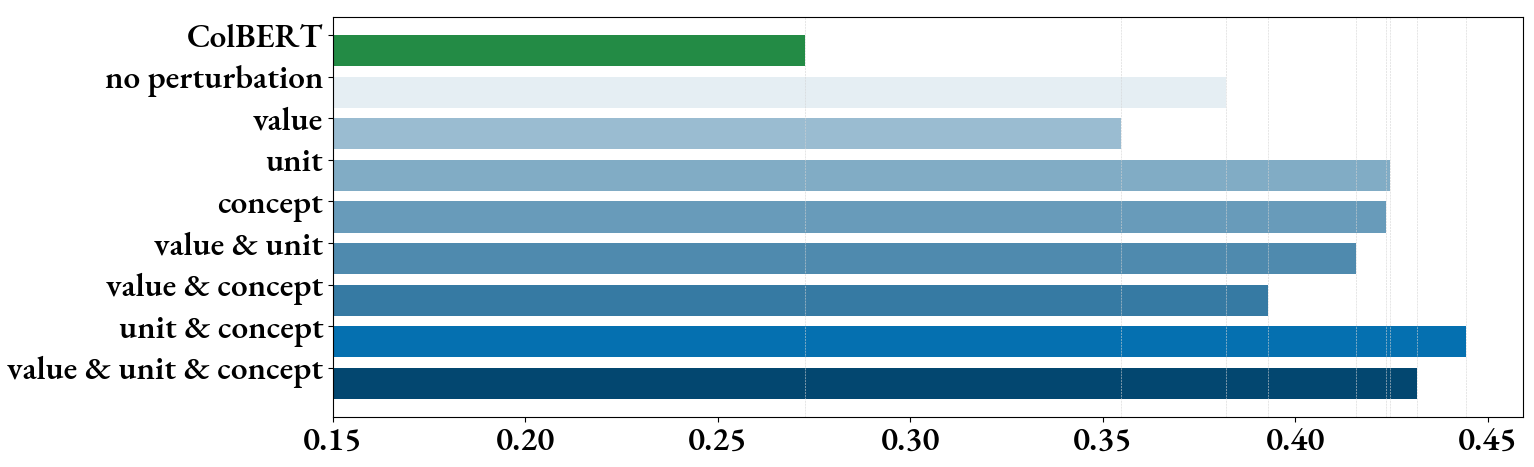}   \caption{ColBERT$_{ft}$}
   \label{fig:colbert_ab} 
\end{subfigure}

\begin{subfigure}[b]{\linewidth}
\includegraphics[width=\linewidth, height=0.29\linewidth]{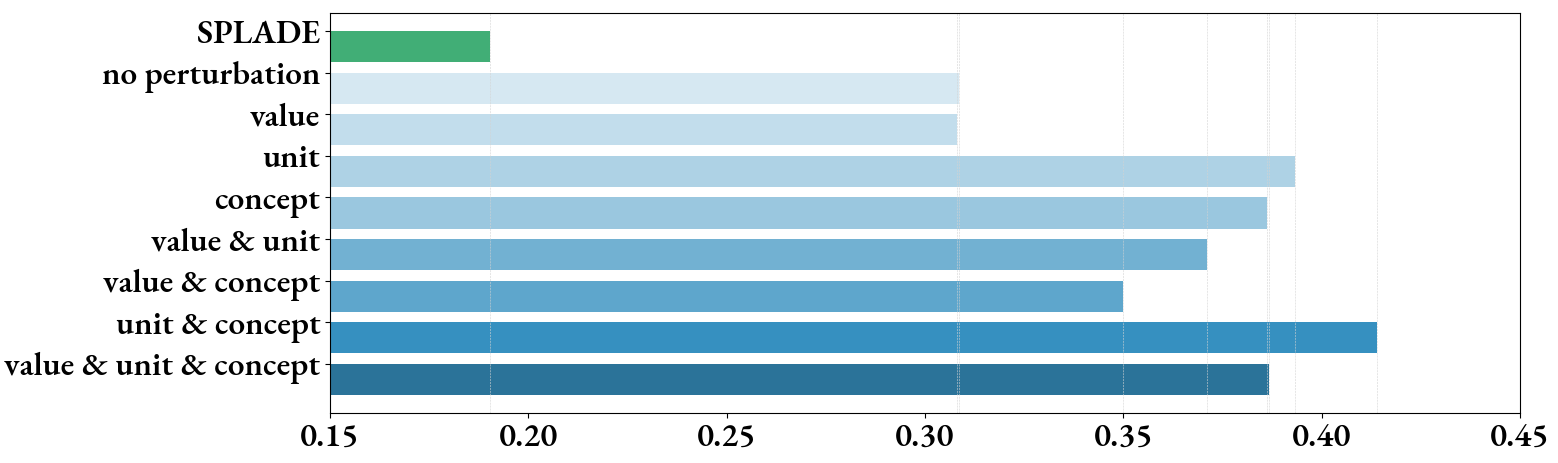}
   \caption{SPLADE$_{ft}$}
   \label{fig:splade_ab}
\end{subfigure}

\caption[]{Ablation study on different augmentation methods, where \emph{value} and \emph{unit}, refer to value and unit permutation and \emph{concept} refers to concept expansion. }
\end{figure}

To check the effect of different augmentation strategies, we perform an ablation study, by fine-tuning the neural models on data generated using a combination of different strategies. 
The main points of variabilities are concept expansion in the query generation process and value and unit permutation for sample generation. 
The results for ColBERT$_{ft}$ and SPLADE$_{ft}$ on FinQuant dataset are demonstrated in Figures~\ref{fig:colbert_ab} and ~\ref{fig:splade_ab}, respectively. 
\emph{no perturbation} refers to the case where no data augmentation was applied and only the positive and negative samples from the original sampling are used for training.

\noindent 
An interesting trend is the detrimental effect of value permutation. 
The value permutation on its own enhances the performance of the base model.
However, as soon as it is accompanied by other augmentation methods the performance degrades slightly. 
The best combination for both the SPLADE and ColBERT model is unit permutation and concept expansion, both of these augmentation techniques on their own also provide a larger boost in comparison to value permutation. 
To this end, the variant of the models presented for evaluation as ColBERT$_{ft}$ and SPLADE$_{ft}$ are trained on unit permutation and concept expansion. 
We find this behaviour rather surprising and counter-intuitive.  
Usually, the performance of neural models increases with the amount of data presented for a given task, however, perturbing the values does not seem to enhance the performance as expected. 
This can be related to the internal representation of the neural models, which is hindering their ability to correctly learn quantity semantics. 
In future work, we aim to test the effect of dedicated numerical embedding and language models for this task. 

\end{document}